\documentclass[12pt]{article}
\usepackage{graphicx}

\newcommand\pubnumber{DPF2015-61}
\newcommand\pubdate{\today}

\def\tamu{Department of Physics and Astronomy\\
Texas A\&M University, College Station, US}
\def\tamuqatar{Department of Physics and Astronomy\\
Texas A\&M University at Qatar, Doha, Qatar}
\def\supporttamu{\footnote{Work supported by US Department of Energy (grant DE-FG02-13ER42020)}}
\def\supporttamuqatar{\footnote{Work supported by Qatar National Research Fund (grant NPRP 5-464-1-080)}}

\def\Title#1{\begin{center} {\Large #1 } \end{center}}
\def\Author#1{\begin{center}{ \sc #1} \end{center}}
\def\Address#1{\begin{center}{ \it #1} \end{center}}

\newcommand\pubblock{\rightline{\begin{tabular}{l} \pubnumber\\
         \pubdate  \end{tabular}}}
\newenvironment{Abstract}{\begin{quotation}  }{\end{quotation}}
\newenvironment{Presented}{\begin{quotation} \begin{center} 
             PRESENTED AT\end{center}\bigskip 
      \begin{center}\begin{large}}{\end{large}\end{center} \end{quotation}}
\def\Acknowledgments{\bigskip  \bigskip \begin{center} \begin{large}
             \bf ACKNOWLEDGMENTS \end{large}\end{center}}

\def\beq{\begin{equation}}
\def\eeq#1{\label{#1}\end{equation}}
\def\eeqn{\end{equation}}

\def\beqa{\begin{eqnarray}}
\def\eeqa#1{\label{#1}\end{eqnarray}}
\def\eeqan{\end{eqnarray}}

\let\bar=\overbar

\def\Dslash{\not{\hbox{\kern-4pt $D$}}}
\def\dslash{\not{\hbox{\kern-2pt $\del$}}}

\def\msb{{\bar{\ssstyle M \kern -1pt S}}}

\usepackage{lineno}
\usepackage{subcaption}
\begin{document}
\begin{titlepage}
\pubblock

\vfill
\Title{Search for non Standard Model Higgs boson decays in events with displaced muon-jets}
\vfill
\Author{{Sven Dildick, Teruki Kamon, Slava Krutelyov, Yuriy Pakhotin, Anthony Rose, Alexei Safonov, Aysen Tatarinov\supporttamu}}
\Address{\tamu}
\Author{{Othmane Bouhali, Alfredo Martin Castaneda Hernandez\supporttamuqatar}}
\Address{\tamuqatar}
\vfill
{
\centering
\large
\hfill On behalf of the CMS Collaboration \hfill 
}
\vfill
\begin{Abstract}
New light bosons that couple weakly to the standard model (SM) particles are predicted in various extensions of the standard model (BSM). Examples include supersymmetric (SUSY) theories with extended Higgs sectors or with a hidden valleys (dark SUSY). In these models the light bosons can be produced directly in the decay of a Higgs boson, or as part of the decay chain of SUSY particles. Depending on the branching fraction, the exotic decays of the SM-Higgs can be undetected in standard analysis techniques or due to its modified production cross section of the Higgs bosons at the LHC. Therefore, direct searches for non-SM decays of the Higgs boson are the fastest way to understand the nature of the Higgs boson. Either it will confirm its SM character, or it will rule out a whole array of BSM scenarios. We present status of the search at CMS for non-SM Higgs boson decays in events with displaced muon-jets.
\end{Abstract}
\vfill
\begin{Presented}
DPF 2015\\
The Meeting of the American Physical Society\\
Division of Particles and Fields\\
Ann Arbor, Michigan, August 4--8, 2015\\
\end{Presented}
\vfill
\end{titlepage}
\def\thefootnote{\fnsymbol{footnote}}
\setcounter{footnote}{0}

\section{Introduction}
The discovery of a scalar boson consistent with the standard model (SM) Higgs boson in 2012 was a milestone in particle physics \cite{cmsHiggs, atlasHiggs}. The discovery concluded a search that lasted nearly 50 years.
The question still remains whether the long sought particle is in fact the SM Higgs boson. To answer this question, continued precision measurements of its properties must be carried out. In particular, the total SM branching ratio must be calculated with high precision. This could take several years of data taking. The current 95\% confidence level (C.L.) upper limit on the non SM branching ratio is 32\% \cite{Khachatryan2014jba}.
An alternative way is to perform a direct search. Either an observation of a non SM Higgs is made or a wide range of beyond the SM scenarios (BSM) is excluded

This paper describes a search for a non SM Higgs boson which decays to new light bosons $a$ \cite{CMS-HIG-13-010}. The light bosons themselves decay to 2 muons. In addition, one or more particles ($X$) can be produced in the decay chain that escape detection. The full Higgs boson decay chain is thus $H \rightarrow 2a + X \rightarrow 4\mu + X$. This search is optimized to discover light bosons with masses between the dimuon mass 
and the ditau mass 
(between $\sim 0.2~\mathrm{GeV}$ and $\sim3~\mathrm{GeV}$) 
The analysis is designed in a such a way that the results are model independent. This allows an interpretation in many different models with boosted dimuons in the event topology. The results are interpreted in two BSM scenarios, a dark SUSY and an next-to-minimal SUSY (NMSSM) scenario. 

The dark SUSY model is motivated by the observation of rising positron fraction cosmic rays up to 200 GeV cosmic ray energy \cite{ams}. The increase in positron fraction can be explained by annihilating dark matter particles. A $U(1)$ symmetry is modeled on top of SUSY, which is broken at the electroweak scale. The Higgs boson decays to SUSY neutralinos, which subsequently decay to dark neutralinos and neutral dark bosons. Since no positron excess is seen in cosmic rays, the dark boson mass is restricted below $2m_p$. The dark boson undergoes kinetic mixing with the SM photon, so that it can decay to light quarks and leptons. In particular when the mass is small, a large branching ratio to dimuons decays is expected.

New light bosons can also be explained within the framework of NMSSM. NMSSM has an extended Higgs sector with 7 Higgs bosons, 3 neutral CP-even states $h_{1,2,3}$, 2 neutral CP-odd states $a_{1,2}$ and 2 charged Higgs states $H^\pm$. The Higgs like scalar boson can be the lightest or the second lightest $h_1$ or $h_2$. Both $h_1$ and $h_2$ can then decay to new light bosons $a_1$. Since $a_1$ has a largely singlet nature it couples to the SM particles. In particular when the $a_1$ mass is between the dimuon mass and the ditau mass, one can expect a large branch ratio to muons \cite{radovan}. 

\section{Analysis method}
The analysis is carried out on data taken with the Compact Muon Solenoid (CMS) at the Large Hadron Collider \cite{Chatrchyan2008aa}. The CMS detector is designed to detect and reconstruct muons with excellent	precision. Information from the tracking system and muon system is used in this analysis. 

This analysis is a follow-up of two previous searches which were conducted at $\sqrt{s} = 7~\mathrm{TeV}$ \cite{CMS-EXO-11-013, CMS-EXO-12-012}. The results for this analysis ($\sqrt{s} = 8~\mathrm{TeV}$) were submitted to Phys. Lett. B. \cite{CMS-HIG-13-010}. Among the results for the $7~\mathrm{TeV}$ analysis included limits on the production cross section times branching ratio squared $\sigma(pp \rightarrow h_{1,2} \rightarrow 2a) \times \mathcal{B}^2 (a\rightarrow 2\mu)$. For prompt dimuons, dark photon masses between 2 muon masses and $2$~GeV were excluded. New in this analysis is the addition of muon displacement, which is added at generation level for the Mont Carlo samples. Displacements are sampled according to an exponential with mean value $0$~mm to $5$~mm. The dark photon is related to the kinetic mixing $\varepsilon_ \mathrm{kin}$ through \cite{Ruderman}. Therefore, once can set limits in a 2D plane ($\varepsilon_ \mathrm{kin}$ vs $m_{\gamma_D}$) \cite{Aad:2014yea}. 

\section{Event selection}
This analysis requires at least 4 muons. Online two muons are selected with a transverse momentum, $p_\mathrm{T}$, of $17$ and $8$ GeV. The offline selection requires 4 muons with $p_\mathrm{T} > 8~\mathrm{GeV}$ and pseudorapidity $|\eta|<2.4 $. At least 1 muon is required to be in the barrel region ($|\eta|<0.9 $) with $p_\mathrm{T} > 17~\mathrm{GeV}$. Muons that are close-by are clustered into pairs of dimuons, based on the vertex probability and the dimuon invariant mass. Exactly 2 dimuons are required in each event. There is however no limit on the number of additional unpaired muons. The dimuons are also required to have the same production origin. This translates to a cut on the vertex parameter $\Delta z < 0.1~\mathrm{cm}$. Finally, the dimuons must have compatible masses, assuming that they are produced from the same kind of dark photons. A diagonal mass corridor is thus constructed in the dimuon-dimuon mass plane. Light SM resonances ($\rho, \omega, \phi, J/\psi$) are used to study the dimuon mass resolution. The dimuons are required to have a mass difference less than $5$ times the mass resolution. 

\section{Displaced muons and pixel hit recovery}
New in this analysis is the displacement of the muons. Muons are produced at a certain distance from the primary vertex. Because of this, the muons will leave fewer hits in the tracker. This results in a much reduced trigger efficiency for muons with a large displacement. To keep the analysis model-independent, a fiducial region in the CMS detector is constructed. Both dimuons are required to have at least 1 hit in the first layer of the barrel pixel or endcap pixel detector, corresponding to $L_{xy} <4.4~\mathrm{cm}$ and $|L_z| < 34.5~\mathrm{cm}$ respectively. This ensures a model independent interpretation of the results. Despite the cut on $L_{xy}$, the analysis is still sensitive to proper displacements of $5~\mathrm{mm}$. The fiducial region for the analysis on $13~\mathrm{TeV}$ will be extended to $L_{xy} \sim 20~\mathrm{cm}$ with the improved trigger.

\begin{figure}[htb]
\centering
\includegraphics[width=.5\textwidth]{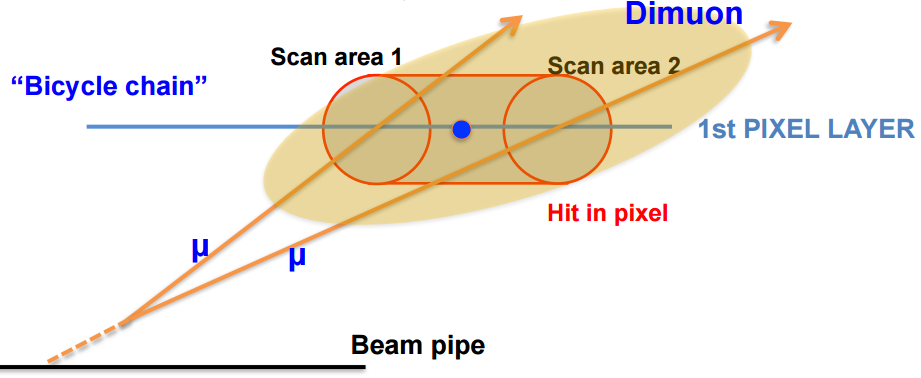}
\caption{Diagram showing the pixel-hit-recovery mechanism developed for this analysis.}
\label{fig:pixel_hit_recovery_mechanism}
\end{figure}

An issue arises when displaced muons are reconstructed in CMS. Namely, when the two muons are highly-parallel and spatially close, the hits in the tracker system can end up being clustered into a single hit, which is not assigned to either muon. The dimuon will therefore fail selection. To recover these dimuons, a recovery technique was developed for this analysis. Muon trajectories are extrapolated to the first layer of the pixel detecor. All nearby pixel hits are collected in a scan area around the two muons and in between them. If a hit was found in the scan area, the dimuon is recovered. This is shown in figure \ref{fig:pixel_hit_recovery_mechanism}.

\section{Results and interpretation}
\begin{figure}[htb!]
   \begin{subfigure}[t]{0.5\textwidth}
        \centering
\includegraphics[width=\textwidth]{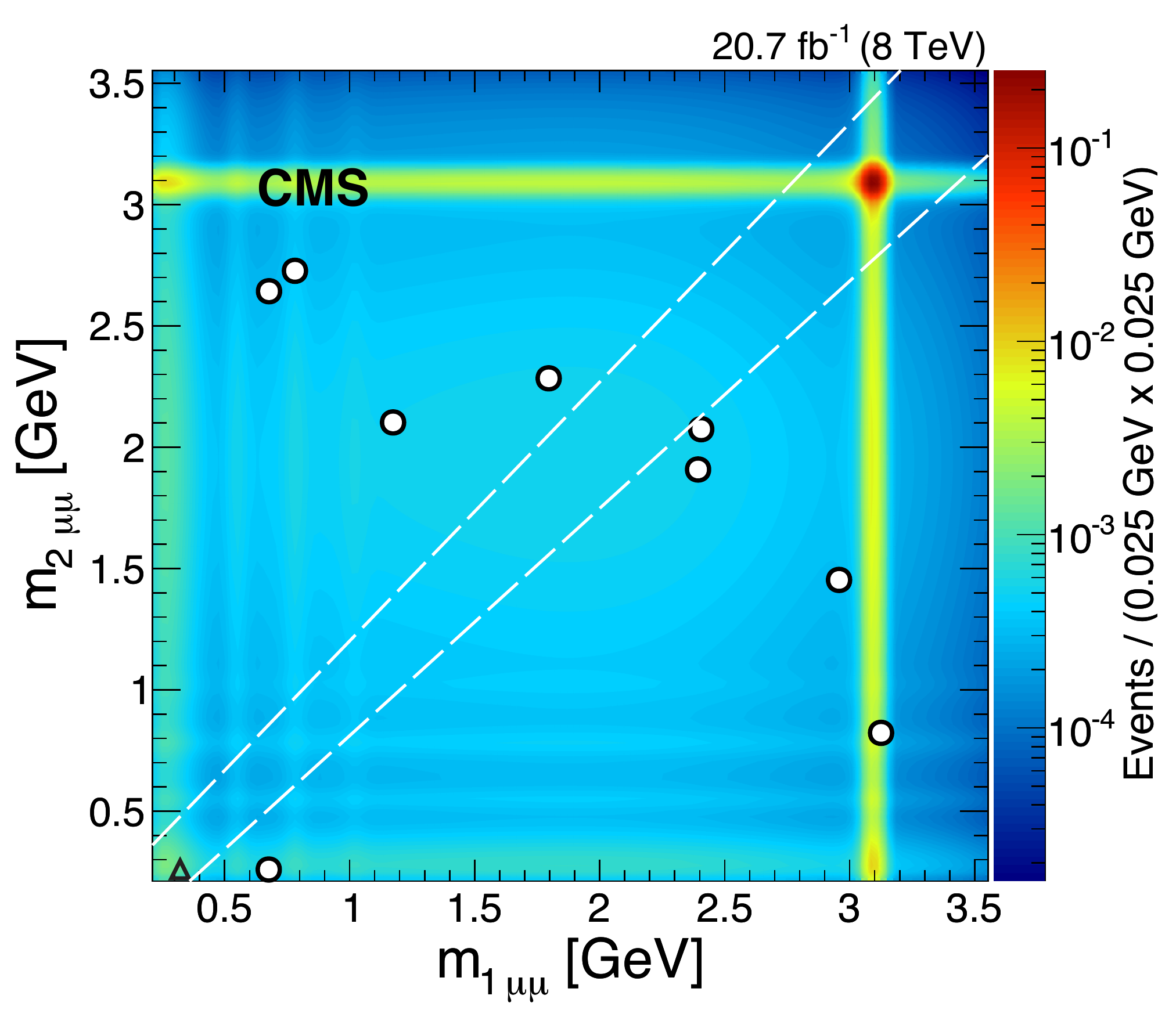}
        \caption{}
        \label{fig:template2D_signal_and_background_m1_vs_m2_v5}
    \end{subfigure}%
    \hfill
   \begin{subfigure}[t]{0.5\textwidth}
        \centering
\includegraphics[width=.86\textwidth]{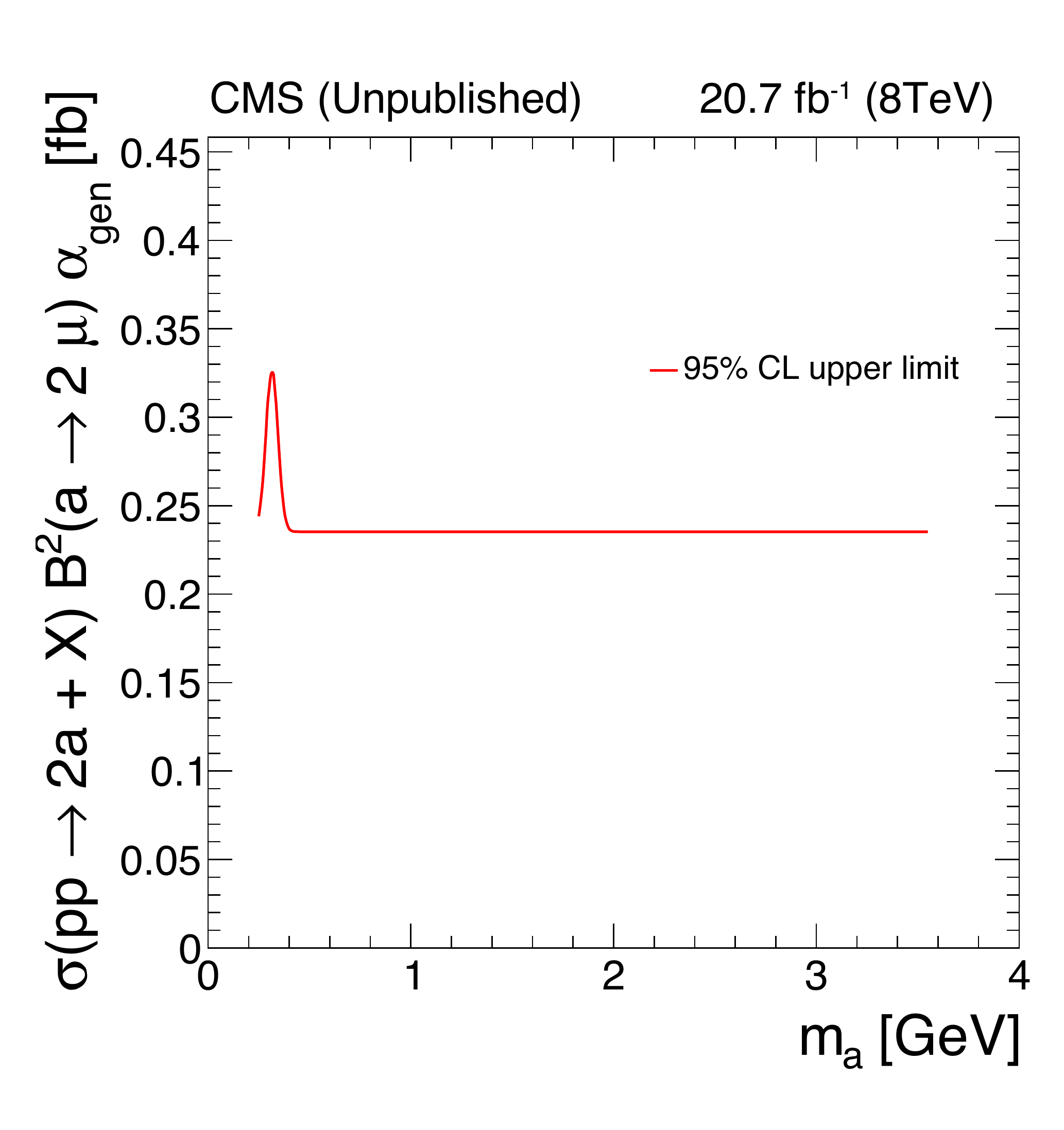}
        \caption{}
        \label{fig:modelIndLimit_vs_ma_2012}
    \end{subfigure}%
\caption{Left: Dimuon mass template for the signal event (green triangle) events that pass all selections, except the dimuon mass constraint (white circles). Right: Model independent 95\% C.L. upper limit on $\sigma(pp \rightarrow 2a) \times \mathcal{B}^2 (a\rightarrow 2\mu) \times \alpha_\mathrm{GEN}$ versus $m_a$, where $\alpha_\mathrm{GEN}$ is the generator level acceptance, which includes the fiducial cut ($L_{xy} <4.4~\mathrm{cm}$ and $|L_z| < 34.5~\mathrm{cm}$).}
\end{figure}

\begin{figure}[htb!]
   \begin{subfigure}[t]{0.5\textwidth}
        \centering
\includegraphics[width=\textwidth]{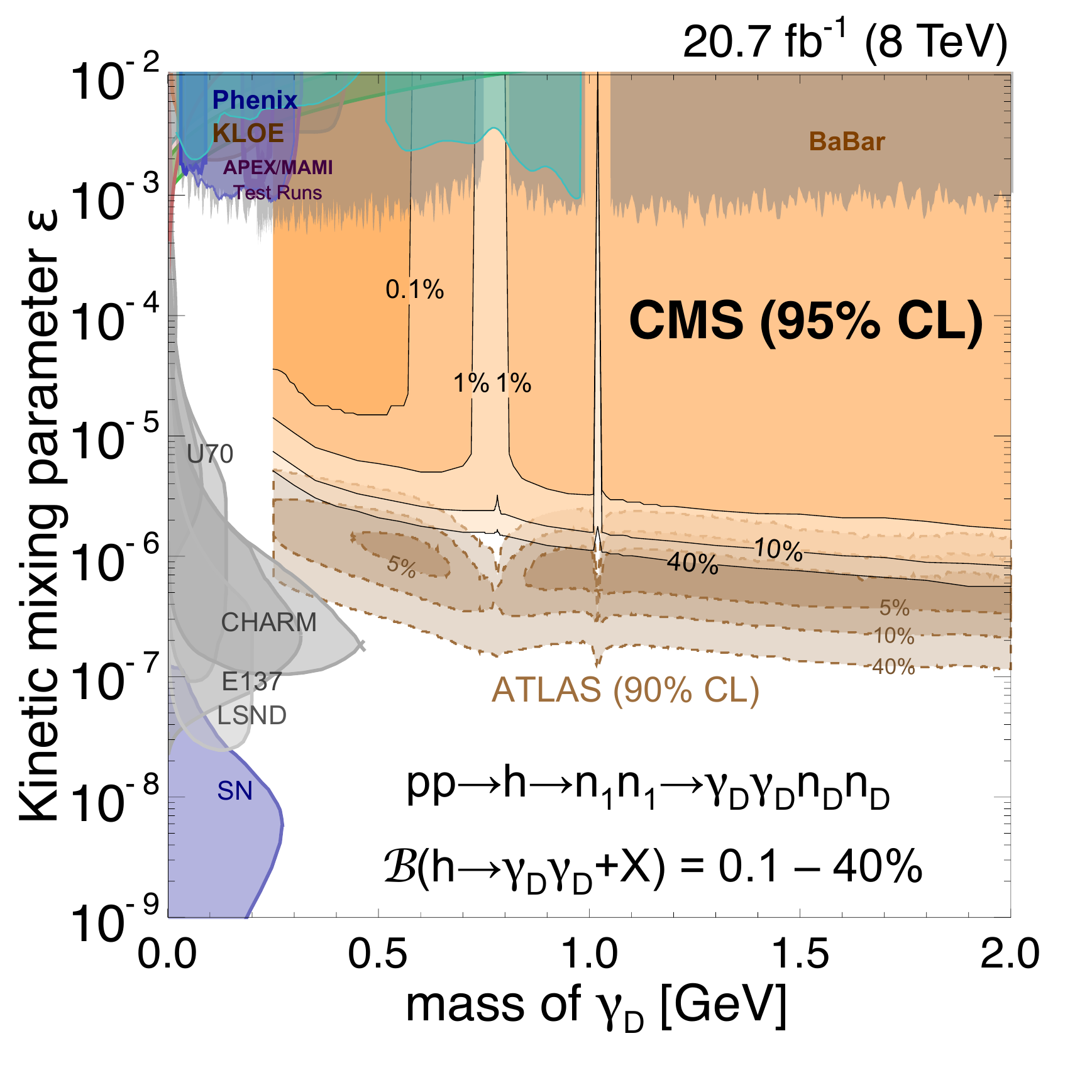}
        \caption{}
        \label{fig:Limit_Eps_mass_v6}
    \end{subfigure}%
    \hfill
   \begin{subfigure}[t]{0.5\textwidth}
        \centering
\includegraphics[width=\textwidth]{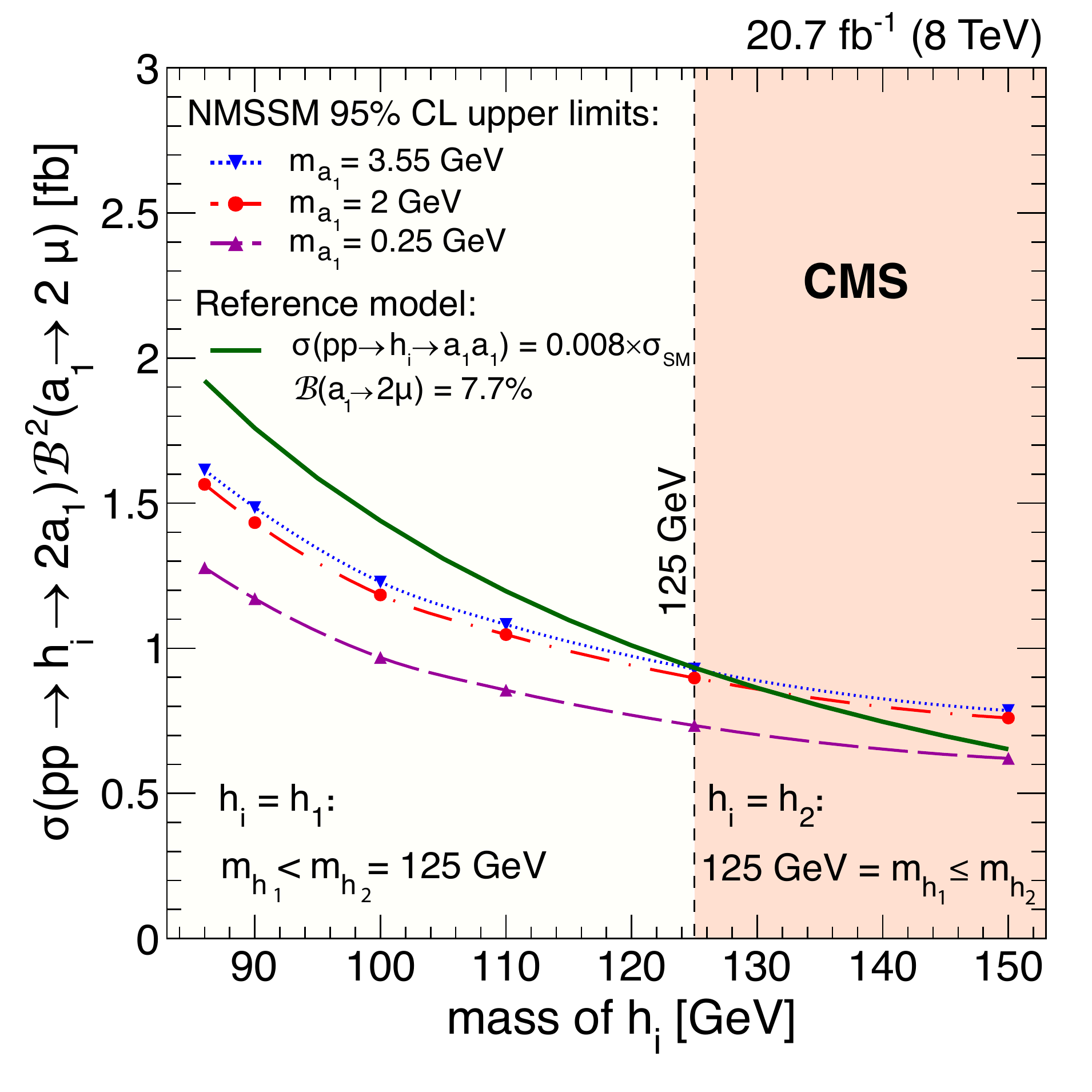}
        \caption{}
        \label{fig:CSxBR_NMSSM_vs_mh_2012_v6}
    \end{subfigure}%
\caption{Left: 95\% C.L. upper limits (black solid curves) on $\sigma(pp \rightarrow h \rightarrow 2\gamma_D + X) \times \mathcal{B}^2 (\gamma_D \rightarrow 2\mu)$ in the plane of the kinetic mixing parameter $\varepsilon_\mathrm{kin}$ and the dark photon mass. The colored contours represent different values of $\mathcal{B} (\gamma_D \rightarrow 2\mu)$ in the range 0.1\% to 40\%. Exclusion limits from other experiments are shown as well. The rapidly varying C.L. as function of $\gamma_D$ is caused by the strong variations of $\mathcal{B} (\gamma_D \rightarrow 2\mu)$. The results from this analysis are complementary to the ATLAS analysis \cite{Aad:2014yea}. Right: 95\% C.L. upper limits in the range $86 < m_{h1} < 125~\mathrm{GeV}$ and $125 < m_{h2} < 150~\mathrm{GeV}$ on $\sigma(pp \rightarrow h_{1/2} \rightarrow 2a_1) \times \mathcal{B}^2 (a_1 \rightarrow 2\mu)$ for $a_1$ boson masses $0.25~\mathrm{GeV}$ (dashed curve), $2~\mathrm{GeV}$ (dashed-dotted curve) and $3.55~\mathrm{GeV}$ (dotted curve). This is compared to the predicted rate from a simplified model (solid curve) with an invisible BSM fraction of 0.8\%. This yields an upper limit on the branching fraction $\mathcal{B}^2 (a_1 \rightarrow 2\mu)$ of 7.7\%.}
\end{figure}

The most significant backgrounds for this analysis are b\={b} decays into a pair of muons, via semi-leptonic decay of a b-quark and the daugher c-quark, or via light hadronic resonances ($\rho, \omega, \phi, J/\psi$), $2.0 \pm 0.7$ events. A secondary background is the electroweak production of dimuons ($pp \rightarrow Z/\gamma^* \rightarrow 4\mu$), $0.15 \pm 0.03$ events. Finally, there is also a minor background from prompt $J/\psi$ production, $0.05 \pm 0.03$ events. After all selections, $2.2 \pm 0.7$ background events were selected. 
After the full event selection only one event survives in data.  This is consistent with the number of background events. Figure  \ref{fig:template2D_signal_and_background_m1_vs_m2_v5} shows shows the selected data event (green triangle) and the events that pass all selections, except the dimuon mass constraint (white circles).
 A model independent limit can be set on $\sigma(pp \rightarrow h_{1,2} \rightarrow 2a) \times \mathcal{B}^2 (a\rightarrow 2\mu)$. The results are shown in figure \ref{fig:modelIndLimit_vs_ma_2012}. 
These results can also be interpreted interpreted in the two benchmark scenarios. Figure \ref{fig:Limit_Eps_mass_v6} shows the 95\% C.L. upper limits on the dark SUSY model. Figure \ref{fig:CSxBR_NMSSM_vs_mh_2012_v6} shows the 95\% C.L. upper limits on NMSSM.

\section{Conclusions}
A search for non-SM Higgs decays to new light bosons in events with displaced muon-jets was presented. One event was observed in $20.7~\mathrm{fb}^{-1}$ of CMS data. This is consistent with the SM expectation of $2.2 \pm 0.7$ background events. A 95\% C.L. model independent limit was set on $\sigma(pp  \rightarrow 2a) \times \mathcal{B}^2 (a\rightarrow 2\mu)$. The results are applicable to a whole range of non-SM scenarios. In particular they were interpreted in two benchmark models, dark SUSY and NMSSM. This analysis will be continued in CMS run-2 with an improved trigger, suitable for displaced muons. 

%
%
%

\Acknowledgments
This work has been completed with the support of the US  Department of Energy (grant DE-FG02-13ER42020) and by the Qatar National Research Fund (grant NPRP 5-464-1-080).

\end{document}